\documentclass[preprint2]{aastex}
\usepackage{graphicx}
\shorttitle{spectral slope variation at proton scales}
\shortauthors{Bruno, Trenchi and Telloni}
\begin{document}

\title{Spectral Slope Variation at Proton Scales from Fast to Slow Solar Wind}

\author{Bruno\altaffilmark{1}\footnotemark[3], R., Trenchi\altaffilmark{1}\footnotemark[3], L. and Telloni\altaffilmark{2}\footnotemark[3], D.}
\altaffiltext{1}{INAF-IAPS, Istituto di Astrofisica e Planetologia Spaziali, Rome, Italy, Via del Fosso del Cavaliere 100, 00133 Roma, Italy}
\altaffiltext{2}{INAF-OATO, Osservatorio Astronomico di Torino, Via dell' Osservatorio 20, 10025 Pino Torinese, Italy}

\email{roberto.bruno@iaps.inaf.it}

\begin{abstract}
We investigated the behavior of the spectral slope of interplanetary magnetic field fluctuations at proton scales for selected high resolution time intervals from WIND and MESSENGER spacecraft at $1$ AU and $0.56$ AU, respectively. The analysis was performed within the profile of high speed streams, moving from fast to slow wind regions. The spectral slope showed a large variability between $-3.75$ and $-1.75$ and a robust tendency for this parameter to be steeper within the trailing edge where the speed is higher and to be flatter within the subsequent slower wind, following a gradual transition between these two states. The value of the spectral index seems to depend firmly on the power associated to the fluctuations within the inertial range, higher the power steeper the slope.
Our result support previous analyses suggesting that there must be some response of the dissipation mechanism to the level of the energy transfer rate along the inertial range.
\end{abstract}


\keywords{interplanetary medium, magnetic fields, plasmas, solar wind, turbulence, waves}

\maketitle
\section{Introduction}
\footnotetext[3]{These authors contributed equally to this work}
Typical time scales of solar wind plasma and magnetic field fluctuations extend over several decades, from the Sun's rotation period down to the smallest scales of the order of ion and electron gyroperiods. The corresponding power density spectrum is characterized by at least three different frequency regions. The lowest frequency range, corresponding to fluctuations of the large scales containing energy, is characterized by a spectral index of the kind $f^{-1}$ \citep{matthaeus1986, dmitruk2004}  whose origin is still debated. A spectral break separates this range from a typical turbulence spectrum as firstly shown by \cite{coleman1968}. In particular, \cite{salem2000}, \cite{podesta2007} and \cite{salem2009} found that the power law exponents of velocity and magnetic field fluctuations, at odds with expectations, often have values near the Iroshnikov-Kraichnan scaling $-3/2$ and the Kolmogorov scaling $-5/3$, respectively. However, as remarked by \cite{roberts2007}, Voyager observations of the velocity spectrum have demonstrated a likely asymptotic state in which the spectrum steepens towards a spectral index of $-5/3$, finally matching the magnetic spectrum and the theoretical expectation of Kolmogorov turbulence. The spectral break cited above was found to shift to lower and lower frequencies with increasing the radial distance from the sun, that is with increasing the age of the turbulence \citep{brunocarbone2013} suggesting that larger and larger scales are continuously involved in the turbulent dynamics which transfers energy from larger to smaller scales to be eventually dissipated at kinetic scales. As a matter of fact, around the proton scales, either the proton inertial length or the proton Larmor radius, there is another spectral break beyond which the spectrum generally steepens. This part of the spectrum is commonly called ``dissipation range'', in analogy to hydrodynamics although the nature of this high frequency part of the interplanetary fluctuations is still largely debated \citep{alexandrova2013, brunocarbone2013}. It was recently shown \citep{brunotrenchi2014} that also this break shifts to lower frequencies as the wind expands. In particular, the same authors showed that the radial dependence of the corresponding wavenumber, of the kind $\kappa_b\sim R^{-1.08}$, is in good agreement with that of the wavenumber derived from the linear resonance condition for proton cyclotron damping (\cite{marsch2003}, \cite{gary2004}, \cite{marsch2006} and references therein). Less clear is the value of the spectral index to be associated to this frequency range. As  matter of fact, \cite{smith2006} performed a wide statistical study on the spectral index in the dissipation range using about $900$ intervals of interplanetary magnetic field recorded by ACE spacecraft at $1$ AU.  These authors found that while within the inertial range the distribution of the values of the spectral index was quite narrow and peaked between $-5/3$ and $-3/2$ that corresponding to the dissipation range was quite broader, spanning from $-1$ to $-4$ with a broad peak between $-2$ and $-3$.
They also found that the dissipation range power-law index at 1 AU strongly depends on the overall fluctuation levels of the interplanetary magnetic field and, consequently, on the rate of energy cascade. In fact, \cite{smith2006} estimated the energy cascade rate $\epsilon$ for all the events they studied and found a clear correlation, with the steepest dissipation range spectra associated with the highest cascade rate. In particular, they found that the spectral index varies with the energy cascade rate $\epsilon$ following $\sim -1.05\epsilon^{0.09}$.
These conclusions supported previous results by \cite{leamon1998} who found a positive correlation between the steepness of the spectral index in the dissipation range and the thermal proton temperature, suggesting that steeper dissipation range spectra imply greater heating rates.
\cite{markovskii2006} found that turbulence spectra often have power-law dissipation ranges with an average spectral index of $-3$ and suggested that this fact is a consequence of a marginal state of the instability in the dissipation range. However, they concluded that their mechanism, acting together with the Landau damping, would produce an entire range of spectral indices, not just $-3$, in better agreement with the observations.

On the other hand, a different view \citep{biskamp1996, ghosh1996, stawicki2001, galtier2007} suggests that beyond the spectral break another turbulent cascade develops. \cite{alexandrova2008}, based on results obtained studying Cluster magnetic field observations in the solar wind, suggested that right beyond the frequency break there is another nonlinear compressible cascade rather than a dissipation range. These authors, introduced a  phenomenological model, based on the compressible Hall MHD and the assumption of kinetic and magnetic energy equipartition, which was able to reproduce the non universality of the spectral slope in the dissipation range simply taking into account the effects of plasma compressibility. In this way, the complete range of variability of the spectral index found by \cite{leamon1998} could be recovered.


However, stimulated by recent results by \cite{brunotrenchi2014}, supporting the fact that a cyclotron-resonant dissipation mechanism which involves the active role of Alfv\'{e}nic fluctuations must participate in the spectral cascade, we looked at the behavior of this high frequency part of the spectrum examining different regions of high speed streams since, moving from fast wind to slow wind across the rarefaction region, both the Alfv\'{e}nicity and the amplitude of the fluctuations greatly change \citep{brunocarbone2013}.

\begin{table}[t]
\begin{center}
\begin{tabular}{ccc} \hline
Interval&s/c& $R(AU)$ \\\hline
2011, 121.500-129.558&WIND&0.99 \\
2011, 175.000-181.558&WIND&0.99 \\
2011, 241.000-245.558&WIND&0.99 \\
2010, 182.000-189.558&WIND&0.99 \\
2010, 182.000-182.038&MESS&0.56 \\
2010, 183.576-183.614&MESS&0.56 \\
2010, 186.628-186.666&MESS&0.56 \\\hline
\end{tabular}
\caption{\label{table} Summary of data intervals used in this analysis.}
\end{center}
\end{table}

\section{Data analysis and results}
In this paper we used observations by WIND at the Lagrangian point $L1$ and MESSENGER in the inner heliosphere. Magnetic field measurements were performed by MFI \citep{lepping1995} onboard WIND at $\sim$11Hz, by MAG \citep{anderson2007} onboard MESSENGER at 20Hz while SWE \citep{ogilvie1995} onboard WIND was used for plasma measurements.
All the time intervals used in this analysis are listed in Table \ref{table}.

We examined several high speed streams observed by WIND, characterized by a smooth and gradual variation of the solar wind speed in the rarefaction region, from fast to slow wind. Within these regions we studied the evolution of the magnetic field fluctuations looking at the total power spectral density (PSD hereafter) derived from the trace of the spectral matrix obtained using a Fast Fourier Transform. Leakage effects were mitigated by a Hanning windowing and, a $33$ points moving average was applied to obtain the spectral estimates.

Figure \ref{figure01}a shows the speed profiles for three of these streams observed by WIND in 2011. The gray shading highlights the regions of interest. Each of them was divided in adjacent sub-intervals of $2^{19}$ points, corresponding to approximately $13$ hours, and for each of them we computed the total PSD of the magnetic field fluctuations.  For sake of clarity, we show only some of these spectra in panels $b$, $c$ and $d$, respectively.
Different colors refer to different time intervals whose starting time and the corresponding wind speed are listed in each panel.

As expected, the PSD is generally higher within the high speed  wind and gradually decreases in the rarefaction region. Moreover, within the selected intervals, plasma is hotter and fluctuations have a stronger Alfv\'{e}nic character dominated by outward modes where the wind speed is higher (not shown).

While at lower frequencies all the spectra have a similar slope, close to $-5/3$ typical of the Kolmogorov scaling, at higher frequencies, above the spectral break observed around $0.3 - 0.4$ Hz, we find a large variability \citep{leamon1998, smith2006}. Steeper spectral slopes are generally observed in high speed wind and are associated with higher PSD in the inertial range. On the contrary, going  towards the slow wind the spectral slope gradually decreases up to a quasi-disappearance of the spectral break.

To check whether this behavior was consistent also with observations in the inner heliosphere,
we analyzed MESSENGER data during the radial alignment  with WIND occurred in July 2010, when Messenger was at $0.56$ AU from the Sun \citep{brunotrenchi2014}. During this alignment,  WIND observed a transition from fast to slow wind (Figure \ref{figure02}a) between June the $30^{th}$ and July the $8^{th}$. The radial alignment between the two spacecraft allowed us to identify the corresponding time interval in MESSENGER's magnetic field data, evaluating the transit time from one s/c to the other on the basis of the wind speed measured by WIND. The possibility to identify in MESSENGER's data similar large scale magnetic field features observed in WIND's data made us confident on the data selection of MESSENGER which observed the same transition from fast to slow solar wind approximately one day before \citep{brunotrenchi2014}.

The methodology used to study this event of WIND is the same described above, the analyzed time interval is shown in Figure \ref{figure02}a and several spectra computed within this time interval are shown in panel $b$. Also in this case the analysis confirms that the spectral slope above the frequency break is strongly related to the wind speed and most likely depends on the power associated with fluctuations within the inertial range.

In case of MESSENGER, the high-resolution ($20$ Hz) magnetic field data were available only for short periods. With this limitation, we evaluated spectra during the three time intervals of $2^{16}$ points listed in Table \ref{table} and indicated by the yellow shadings in Figure \ref{figure02}a.  The corresponding spectra are shown in panel $b$. The average speed values corresponding to these PSDs are deduced from those measured by WIND. MESSENGER strikingly confirms the spectral steepening observed by WIND at $1$ AU beyond the spectral break that for MESSENGER is shifted to higher frequency \citep{brunotrenchi2014}.


In order to relate in some analytical form the observed spectral slope at ion scales to the PSD observed in the inertial range we report our observations in Figure  \ref{figure03} for all the $56$ spectra examined in this work plotting the spectral slope at ion scales as a function of the normalized power (see below) in the inertial range at 1 AU.  The MESSENGER  observations at 0.56 AU are also included in this figure after extrapolation to $1$ AU of the estimated PSD assuming a WKB like radial dependence \citep{Hollweg1973, zhou1989} as discussed below.

The estimate of the power is obtained as the integral of the PSD in a frequency range chosen within the inertial range. For the WIND spectra this frequency range is from $7\times 10^{-3}$ to $ 10^{-1}$ Hz while, for the MESSENGER spectra, we integrated the PSDs in the $2\times 10^{-2}$ to $2\times 10^{-1}$ Hz range.

At lower frequencies, it was necessary to choose a different limit for the MESSENGER spectra, since these spectra were evaluated within shorter data intervals. Also the upper limit was chosen at higher frequency because of the frequency shift of the spectral break observed in the inner heliosphere \citep{brunotrenchi2014}.

Subsequently, we estimated the power that MESSENGER's spectra should have in the same frequency band of WIND by fitting the PSDs with a power law of the kind $f^{-5/3}$. Afterwards, we extrapolated these values to 1 AU, assuming the standard $R^{-3}$ radial dependence predicted by WKB theory \citep{Hollweg1973}. Having noticed that, at $1$ AU, the power spectra corresponding to the lowest speed are approximately at the same low level regardless of the time interval under consideration, we normalized the values of the integrated PSD to the same lowest power in the inertial range obtained throughout our analysis. This value refers to the low speed wind observed by WIND on July the $8^{th}$, 2010. This normalization was applied just to have a dimensionless parameter on the $X$ axis of the plot in Figure \ref{figure04}.

The spectral slopes in the ``dissipation range'' were obtained through a fitting procedure, having care of not including regions too close to the break point or at higher frequencies where the spectrum flattens out \citep{brunotrenchi2014}.

The dependence of the spectral slopes in the dissipation range on the power level in the corresponding inertial range, shown in Figure \ref{figure03}, is rather robust since the same kind of relationship applies equally well to data points belonging to different time intervals indicated by different colors in the plot.
The best fit was obtained using a power law fit, shown by the continuous black line, of the kind:

\begin{equation}\label{fit}
     q=(-4.37\pm0.48)+(2.46\pm0.45){w/w_0}^{(-0.30\pm0.10)}
\end{equation}

where $q$ is the spectral index and $w/w_0$ indicates the normalization process performed within the inertial range.
It is interesting to notice that equation (\ref{fit}) provides an upper limit for the slope in the dissipation range of 4.37, which is very close to the steepest slopes observed in previous studies \citep{smith2006}. In particular, in Figure 3 of \cite{sahraoui2010}, those authors show a spectral behavior, around the frequency break, remarkably similar to what we found within our fast wind streams. Finally, the dependence we found on the wind speed implied also a dependence on the plasma $\beta$ which varied between $0.7$ and $1.7$ (not shown). Thus, lower values of $\beta$ are associated to faster wind and steeper spectral indexes.

\subsection{Noise due to digitization process}

One should be careful while analyzing fluxgate magnetometer data at frequencies which might be influenced by the ring noise due to digitization process \citep{russell1972, lepping1995, smith1998, alexandrova2013} and, in general, by the sensor noise as shown by \cite{howes2008}. \cite{bennett1948} derived the PSD of a uniformly quantized Gaussian random process.
He showed that a uniform quantizer $q$ characterized by a quantization step $\delta$ produces an average distortion $\Delta(q)\sim \delta^2/12$. If we evenly spread this digital noise across the frequency band $0$ to the Nyquist frequency $f_N$, the power density level expected for the digitization noise $W(\delta)$ would be: $W(\delta)\sim\delta^2/(12f_N)$.
Thus, using this expression and knowing the quantization step $\delta$ we can estimate the spectral level due to quantization in case of Messenger and WIND, respectively.

In case of MESSENGER (Figure \ref{figure02}) the spectral flattening above roughly $3$ Hz gives an average spectral density higher than the lowest detectable power associated with the digitization step size of $0.047$ nT \citep{anderson2007} which would be around $1.8\times10^{-5} \textrm{nT}^2/\textrm{Hz}$. However, as shown by \cite{anderson2007} the lower detection limit $W(\delta)$ is not a comprehensive measure of the entire high-frequency noise due to digitization. This level was estimated \citep{anderson2007} to be around $2.5\times10^{-4} \textrm{nT}^2/\textrm{Hz}$ and it is clearly visible also in our spectra relative to Messenger (see Figure \ref{figure02}) at frequencies larger than $2-3 \textrm{Hz}$. Anyhow, being at much lower level, cannot determine the less steep slope observed right beyond the spectral break already visible for the second spectrum from the top (light blue color).

\begin{figure}\abovecaptionskip 1 mm \belowcaptionskip 1 mm
\includegraphics[width=6cm]{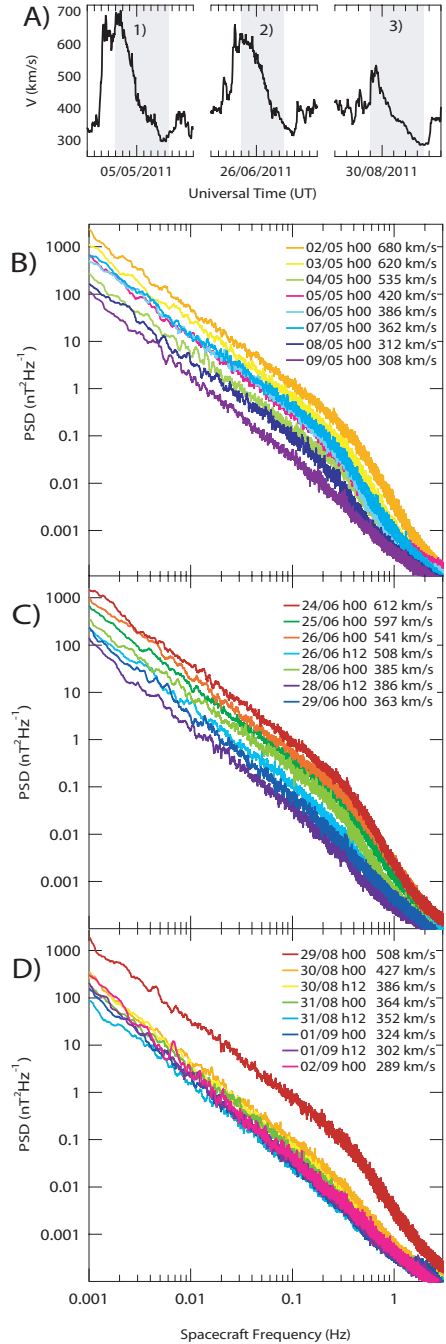}
\caption{Panel a: wind speed profile of three different high speed streams observed by WIND at $1$ AU. The following panels b, c and d show a selection of magnetic field spectra obtained along the speed profile of fast streams $1)$, $2)$ and $3)$, respectively, within the time intervals indicated by the shaded areas.}\label{figure01}
\end{figure}

\begin{figure}\abovecaptionskip 1 mm \belowcaptionskip 1 mm
\includegraphics[width=6.5cm]{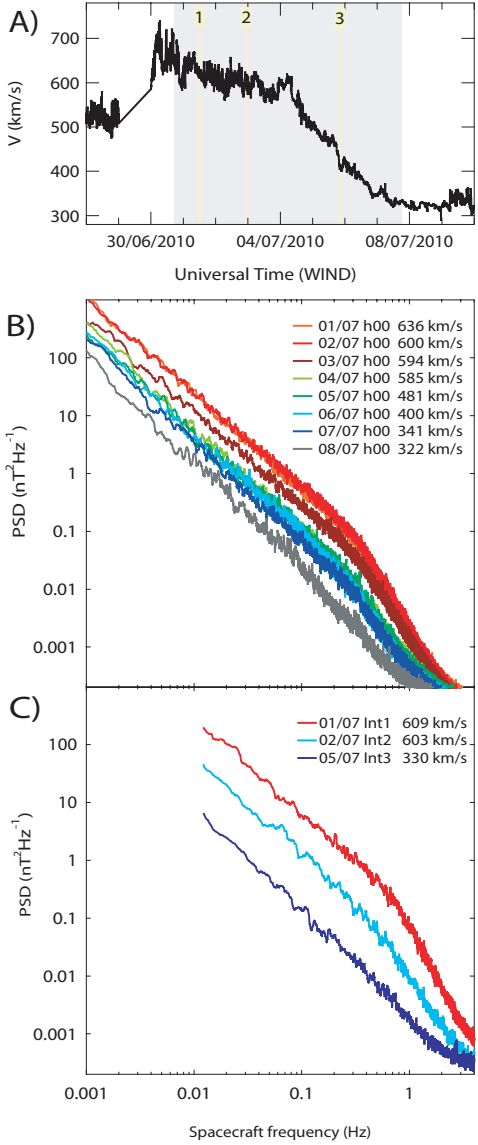}
\caption{In the same format as of Figure \ref{figure01}, panel a shows the speed profile of a fast stream observed by WIND at the beginning of July 2010. Panel b shows some of the spectra computed for several time intervals within the shaded area shown in panel a. Panel c refers to MESSENGER's spectra computed within the three intervals indicated by the yellow shading in panel a.}\label{figure02}
\end{figure}

\begin{figure}\abovecaptionskip 1 mm \belowcaptionskip 1 mm
\includegraphics[width=6.5cm]{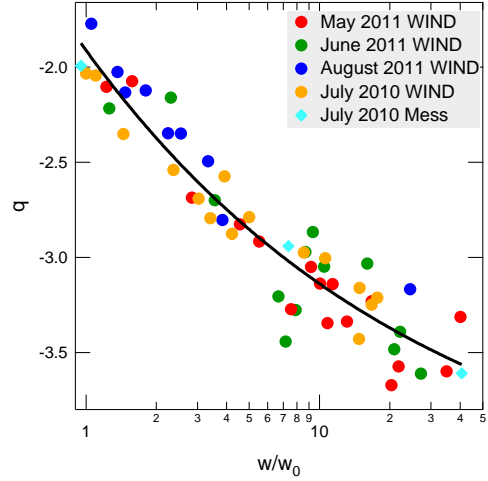}
\caption{Distribution of normalized values of integrated PSD in the inertial range (see text for details) versus values of the spectral index in the dissipation range at proton scales. Different symbols refer to different time intervals studied within different high speed streams observed at $1$ AU by WIND and at $0.56$ AU by MESSENGER.}\label{figure03}
\end{figure}

Similar evaluations would suggest that the lowest detectable power associated with the digitization step size in case of WIND \citep{lepping1995} would be around $1.7\times10^{-5}\textrm{nT}^2/\textrm{Hz}$ assuming a digital step of $0.032$ nT. However, also in this case the digitization noise level flattens out the spectra at a level of roughly $5\times10^{-4}\textrm{nT}^2/\textrm{Hz}$ for frequencies beyond $2$ Hz (see Figure \ref{figure01}).

\begin{figure}\abovecaptionskip 1 mm \belowcaptionskip 1 mm
\includegraphics[width=6.5cm]{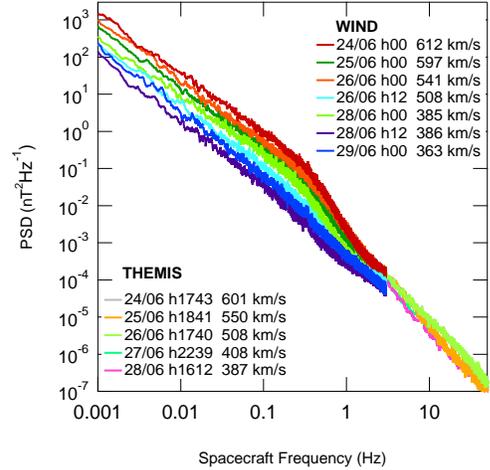}
\caption{WIND spectra shown in Figure \ref{figure01}C are extended into the dissipation range adopting spectra from THEMIS-C search-coil observations at $128$Hz.}
\label{figure04}
\end{figure}

As further proof about the reliability of our spectral estimates at ions scales, we analyzed data from the search-coil magnetometer onboard THEMIS-C \citep{Roux2008} during the same stream of June 2011 observed by WIND and obtained solar wind speed values from the plasma sensor sensor \citep{mcfadden2008}. During part of the duration of this stream, THEMIS-C was in the solar wind not connected to the Earth's bow-shock. These spectra, based on $128$Hz sampling frequency,  are shown, together with WIND's spectra in Figure \ref{figure04} and unravel the behavior of these spectra within the next frequency decade or so. The length of each data sample varied between $2^{13}$ and $2^{15}$ data points and the relative starting time is shown in the same Figure. Frequency spikes, due to the fact that data have not been de-spun, have been removed artificially from the graph leaving unaltered the general behavior of the spectra which satisfactorily matches WIND spectra at $3$Hz and show that the power level, at these scales, is independent on whether the spectrum refers to fast or slow wind. The average value of the spectral index between $3$ and $50$ Hz, for the analyzed periods is $-2.36 \pm 0.11$.

\section{Summary and conclusions}

We investigated the behavior of the spectral slope at proton scales, up to frequencies of a few Hz, beyond the high frequency break separating fluid from  kinetic scales. We used high time resolution interplanetary magnetic field data from WIND and MESSENGER spacecraft at $1$ AU and $0.56$ AU, respectively. Several time intervals were selected whenever long enough samples of high resolution data were available within high speed streams and the following rarefaction regions down to the slow wind. In particular, we avoided to analyze those regions of slow wind characterized by strong compressive phenomena due to the dynamical interaction between fast and slow wind and showing a clear lack of time stationarity. One of these selected events corresponded to a radial alignment between the two spacecraft already studied by \cite{brunotrenchi2014}. We found a large variability of the spectral slope, as already reported in literature \citep{leamon1998, smith2006}, between $-3.75$ and $-1.75$. However, we also found a robust tendency for this parameter to show the steepest spectra within the trailing edge of the fast streams where the speed is higher and the lowest values within the subsequent slow wind, following a gradual transition between these two states. The value of the spectral index seems to depend firmly on the power characterizing the fluctuations within the inertial range, higher the power steeper the slope.
In particular, this slope tends to approach $-5/3$ within the slowest wind and a limiting value of $-4.37\pm0.48$ within the fast wind.

Our result support previous analyses \citep{smith2006} suggesting that there must be some response of the dissipation mechanism related to the energy transfer rate along the inertial range.
Generally, fluctuations within faster wind not only have larger amplitude but are also more Alfv\'{e}nic.
It would be interesting to understand whether the spectral slope in this high frequency range depends solely on the power level of the fluctuations or also on their Alfv\'{e}nic character, i.e. on the crosshelicity. In any case, average values of the spectral index at proton scales, based on statistical studies employing large dataset, would depend on the relative amount of fast and slow wind present in the dataset itself. In this respect, the same analysis performed for different phases of the solar cycle, characterized by a different amount of fast wind in the ecliptic, could produce contradictory results which would merely depend on an unbalanced presence of fast and slow wind.

\section{Acknowledgments}
This research was partially supported by the Agenzia Spaziale Italiana under contract ASI/INAF I/013/12/0  and by the European Community's Seventh Framework Programme (FP7/2007-2013) under grant agreements N. 313038/STORM. Data from WIND, THEMIS-C and MESSENGER were obtained from NASA-CDAWeb and NASA-PDS websites, respectively.



\begin{thebibliography}{10}
\bibitem[Alexandrova et al. (2008)]{alexandrova2008} Alexandrova, O., Carbone, V., Veltri, P., Sorriso-Valvo, L.\ 2008.\ ApJ, 674, 1153-1157.



\bibitem[Alexandrova et al.(2013)]{alexandrova2013} Alexandrova, O., Chen, C.~H.~K., Sorriso-Valvo, L., Horbury, T.~S., Bale, S.~D.\ 2013.\ SSRv 178, 101-139.

\bibitem[Anderson et al.(2007)]{anderson2007} Anderson, B.~J., Acu{\~n}a, M.~H., Lohr, D.~A., Scheifele, J., Raval, A., Korth, H., Slavin, J.~A.\ 2007. SSRv 131, 417-450.






\bibitem[Bennett (1948)]{bennett1948} Bennett, W. \ 1948. Bell System Technical Journal, 27, 446–472.

\bibitem[Biskamp et al.(1996)]{biskamp1996} Biskamp, D., Schwarz, E., Drake, J.~F.\ 1996.\ PhRvL 76, 1264-1267.



\bibitem[Bruno and Carbone (2013)]{brunocarbone2013} Bruno, R., Carbone, V.\ 2013.\ LRSP, 10, 2.

\bibitem[Bruno and Trenchi(2014)]{brunotrenchi2014} Bruno, R., Trenchi, L.\ 2014.\ ApJ, 787, L24.



\bibitem[Coleman (1968)]{coleman1968} Coleman, P.J.\ 1968. ApJ, 153, 371


\bibitem[Dmitruk et al. (2004)]{dmitruk2004} Dmitruk, P., Matthaeus, W.~H., Seenu, N.\ 2004.\ ApJ, 617, 667-679.


\bibitem[Galtier and Buchlin (2007)]{galtier2007} Galtier, S., Buchlin, E.\ 2007.\ ApJ, 656, 560-566.


\bibitem[Gary and Borovsky (2004)]{gary2004} Gary, S.~P., Borovsky, J.~E.\ 2004.\ JGRA, 109, 6105.

\bibitem[Ghosh et al.(1996)]{ghosh1996} Ghosh, S., Siregar, E., Roberts, D.~A., Goldstein, M.~L.\ 1996.\ JGRA 101, 2493-2504.

\bibitem[Hollweg(1973)]{Hollweg1973} Hollweg, J.~V.\ 1973, ApJ, 181, 547


\bibitem[Howes et al.(2008)]{howes2008} Howes, G.~G., Cowley, S.~C., Dorland, W., Hammett, G.~W., Quataert, E., Schekochihin, A.~A.\ 2008.\ JGRA, 113, 5103.


\bibitem[Leamon et al. (1998)]{leamon1998} Leamon, R.~J., Smith, C.~W., Ness, N.~F., Matthaeus, W.~H., Wong, H.~K.\ 1998.\ JGRA, 103, 4775.



\bibitem[Lepping et al.(1995)]{lepping1995} Lepping, R.~P., et al., 1995. SSRv 71, 207-229.


\bibitem[Markovskii et al.(2006)]{markovskii2006} Markovskii, S.~A., Vasquez, B.~J., Smith, C.~W., Hollweg, J.~V.\ 2006.\ ApJ, 639, 1177-1185.



\bibitem[Marsch et al.(2003)]{marsch2003} Marsch, E., Vocks, C., Tu, C.-Y.\ 2003.\ NPG 10, 101-112.

\bibitem[Marsch (2006)]{marsch2006} Marsch, E.\ 2006.\ LRSP, 3, 1.


\bibitem[Matthaeus and Goldstein, 1986]{matthaeus1986} Matthaeus, W.~H., Goldstein, M.~L.\ 1986.\ PhRvL, 57, 495-498.

\bibitem[McFadden et al.(2008)]{mcfadden2008} McFadden, J.~P., Carlson, C.~W., Larson, D., et al.\ 2008, \ SSR, 141, 277

\bibitem[Ogilvie et al.(1995)]{ogilvie1995} Ogilvie, K.~W. et al., 1995. \ SSRv 71, 55-77.


\bibitem[Podesta et al. (2007)]{podesta2007} Podesta, J.~J., Roberts, D.~A., Goldstein, M.~L.\ 2007. \ ApJ, 664, 543-548.

\bibitem[Roberts (2007)]{roberts2007} {Roberts}, D.~A.\ 2007. \ AGU Fall Meeting Abstracts

\bibitem[Roux et al.(2008)]{Roux2008} Roux, A., Le Contel, O., Coillot, C., et al.\ 2008, \ssr, 141, 265

\bibitem[Russell (1972)]{russell1972} {Russell}, C. T., \ 1972. \ Solar Wind, NASA sp-308, 365-374


\bibitem[Sahraoui et al.(2010)]{sahraoui2010} Sahraoui, F.,
Goldstein, M.~L., Belmont, G., Canu, P., Rezeau, L.\ 2010.\  PhRvL 105, 131101.

\bibitem[Salem (2000)]{salem2000} Salem, C.,\ 2000. \ PhD thesis, Univ. Paris VII

\bibitem[Salem et al., (2009)]{salem2009} Salem, C., Mangeney, A., Bale, S.D., Veltri, P.\ 2009. \ ApJ 702, 537–553


\bibitem[Smith et al.(1998)]{smith1998} Smith, C.~W., L'Heureux, J., Ness, N.~F., Acu{\~n}a, M.~H., Burlaga, L.~F., Scheifele, J.\ 1998.\ SSRv 86, 613-632.

\bibitem[Smith et al. (2006)]{smith2006} Smith, C.~W., Hamilton, K., Vasquez, B.~J., Leamon, R.~J.\ 2006.\ ApJ, 645, L85-L88.

\bibitem[Stawicki et al.(2001)]{stawicki2001} Stawicki, O., Gary, S.~P., Li, H.\ 2001.\ JGRA 106, 8273-8282.

\bibitem[Zhou and Matthaeus(1989)]{zhou1989} Zhou, Y., Matthaeus, W.~H.\ 1989.\ GRL 16, 755-758.








\end{thebibliography}
\end{document}